\documentclass[twocolumn,aps,prl,superscriptaddress]{revtex4-2}
\usepackage{amsfonts}
\usepackage{amsmath}
\usepackage{amssymb}
\usepackage{graphicx}
\usepackage{color}
\usepackage{ulem}
\usepackage[colorlinks, urlcolor=cyan, citecolor=blue, linkcolor=magenta]{hyperref}

\begin{document}

\title{Detecting Entanglement via Split Spectroscopy in Many-Body Systems}
\author{Hao-Yue Qi}
\affiliation{Hefei National Research Center for Physical Sciences at the Microscale and
School of Physical Sciences, University of Science and Technology of China,
Hefei 230026, China}
\affiliation{CAS Center for Excellence in Quantum Information and Quantum Physics,
University of Science and Technology of China, Hefei 230026, China}
\author{Wei Zheng}
\email{zw8796@ustc.edu.cn}
\affiliation{Hefei National Research Center for Physical Sciences at the Microscale and
School of Physical Sciences, University of Science and Technology of China,
Hefei 230026, China}
\affiliation{CAS Center for Excellence in Quantum Information and Quantum Physics,
University of Science and Technology of China, Hefei 230026, China}
\affiliation{Hefei National Laboratory, University of Science and Technology of China,
Hefei 230088, China}
\date{\today }

\begin{abstract}
Quantum entanglement is recognized as a fundamental resource in quantum information processing and is essential for understanding quantum many-body physics. However, experimentally detecting entanglement, particularly in many-particle quantum states, remains a significant challenge. Here, we propose split spectroscopy as an experimentally feasible technique for detecting entanglement of eigenstates in quantum many-body systems. We demonstrate the split spectroscopy exhibits a single delta-function peak if and only if the investigated eigenstate is triseparable. Our framework is illustrated using two paradigmatic spin models that undergo quantum phase transitions. Furthermore, we show that the spectral entropy serves as a powerful indicator of quantum phase transitions and captures the scaling behavior of entanglement. Finally, we present an experimental protocol using Rydberg atom arrays.
\end{abstract}

\date{\today}
\maketitle
\textit{Introduction.}
With the advent of quantum information science, entanglement has been recognized as a pivotal resource \cite{Gour@2019.QResource}, enabling supremacy in quantum communications \cite{Tavakoli@2022.QCommu}, quantum computing \cite{Nielsen@2000.QCompu}, and metrology \cite{Treutlein@2018.QMetro,Cappellaro@2017.QSense}. It also enriches our understanding of fundamental physics across various areas. For example, the holographic principle is intimately connected to area law scaling of entanglement entropy in microscopic theories \cite{Bousso@2002}. The scaling behavior of entanglement serves as a diagnostic of the quantum chaos and integrability in many-body systems \cite{Rigol@2018,Srednicki@2019,Vidmar@2022}. Moreover, entanglement plays a crucial role in the study of quantum critical phenomena \cite{Zanardi@2002,Kitaev@2003,Lin@2004,Ortolani@2006,Plenio@2010}.

Given its fundamental importance, numerous entanglement measures \cite{Virmani@2007,Horodecki@2009} have been devised to quantify it, including entanglement entropy \cite{Greiner@2015,Greiner@2016,Cardy@2004,Preskill@2006,Maestro@2018,Melko@2010} and squashed entanglement \cite{Winter@2004,Song@2009,Winter@2014}. However, experimental detection of these quantities requires state tomography, which is rarely possible in many-body systems. While randomized measurements are effective for detecting R\'{e}nyi etropy \cite{Zoller@2018,Roos@2019,Zoller@2019,Vermersch@2020,Ma@2022},  they still demand precise control over individual qubits \cite{Lewenstein@2023}. Entanglement witnesses \cite{Toth@2009,Sarbicki@2014,Tennant@arxiv} present a more feasible way to detect the presence of entanglement, utilizing observables such as magnetic susceptibility \cite{Zeilinger@2006,Brukner@2005} and heat capacity \cite{Brukner@2008,Mitra@2013}. These quantities exceed certain bounds only when the investigated state is entangled. However, values within these bounds do not necessarily indicate a separable state. In practice, a priori information about the state is required to select an appropriate witness, rendering some tasks inefficient. Therefore, it is essential to develop a method that provides a necessary and sufficient criterion for entanglement verification, while remaining compatible with experimentally accessible resources.

Spectroscopic measurement has emerged as an experimentally accessible technique across various platforms, allowing for the characterization of intrinsic properties in many-body systems. In condensed matters, the angle-resolved photoemission spectroscopy (ARPES) \cite{Shen@2021} has been employed to study the pseudogap of high-$T_c$ cuprates. In ultracold atomic gases, radio frequency (RF) spectroscopy \cite{Levin@2009} is used to measure single-particle spectra of Fermi gases \cite{Ketterle@2008,Jin@2008,Zhang@2012,Zwierlein@2012} and to explore confinement phase transitions \cite{Zheng@2024}. Notably, quantum Fisher information (QFI) \cite{Smerzi@2012,Toth@2012,Oberthaler@2014}, a witness of multipartite entanglement, has been shown to be measurable using spectroscopic techniques \cite{Zoller@2016,Tennant@2023,Nagar@2020,Alvarez@2021,Konik@arXiv}.

In this paper, we extend the spectroscopic toolkit by proposing a split spectroscopy which probes the response of a quantum system when perturbed by splitting it into three parties, $A$, $B$, and $M$. Generally, the split spectroscopy provides a necessary and sufficient criterion for entanglement verification of eigenstates in quantum many-body systems. Specifically, we demonstrate the spectrum exhibits a single delta-function peak if and only if the investigated eigenstate is separable with respect to subsystems $A$, $B$ and $M$. To illustrate our framework, we analyze two one-dimensional quantum spin models that exhibit quantum phase transitions. Our numerical results validate the proposed criterion. Furthermore, we show that the spectral entropy of ground states, analogous to squashed entanglement, displays singularities near quantum critical points, thereby
serving as a powerful indicator of quantum phase transitions. Remarkably, the spectral entropy also captures the scaling behavior of entanglement. Finally, we discuss experimental realization of the split spectroscopy using Rydberg atom arrays.

\begin{figure}
  \centering
  \includegraphics[width=\columnwidth]{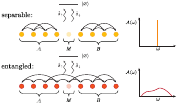}
  \caption{Schematic illustration of the split spectroscopy. A quantum chain is divided into three distinct subsystems $A$, $B$, and $M$. The arrows represent entangling correlations between sites. The split operator $\hat{\mathcal{S}}=\Omega_{\uparrow}\hat{s}_{\uparrow}+\Omega_{\downarrow}\hat{s}_{\downarrow}$ transfers states within subsystem $M$ to an external state $|{\o}\rangle$, thereby splitting the chain into subsystems $A$ and $B$. The spectrum $\mathcal{A}(\omega)$ manifests a singe delta-function peak for separable state with respect to subsystems $A$, $B$ and $M$. In contrast, it displays multiple peaks for entangled states.}\label{fig1}
\end{figure}

\textit{Concepts of split spectroscopy.}
We consider a generic quantum system with open boundary conditions, and divide it into three distinct subsystems $A,B$ and $M$ with sizes $L_A, L_B$ and $L_M$, as illustrated in Fig.\ref{fig1}. The Hamiltonian of the system can be expressed as
\begin{equation}
  \hat{H}=\hat{H}_A+\hat{H}_B+\hat{H}_M+\hat{H}_{AM}+\hat{H}_{BM},
\end{equation}
where $\hat{H}_{X}$ denotes the Hamiltonian for subsystem $X$, and $\hat{H}_{XY}$ represents the coupling terms between subsystems $X$ and $Y$. The Hilbert space of the system is given by $\mathcal{H}=\mathcal{H}_A\otimes\mathcal{H}_M\otimes\mathcal{H}_B$, where $\mathcal{H}_X$ denotes the Hilbert space of subsystem $X$.

Spectroscopic measurements reflect the response of a system to perturbations. The primary distinction among  different spectroscopies is the nature of the perturbation applied to the system. In our approach, the perturbation effectively splits the system.
To formalize this, we extend the Hilbert space $\mathcal{H}_B$ by introducing an auxiliary state $|{\o}\rangle$, which is orthogonal to all states in $\mathcal{H}$ and does not couple with them. The extended Hilbert space is $\mathcal{H}_A\otimes(\mathcal{H}_M\oplus \{|{\o}\rangle\})\otimes\mathcal{H}_B$, in which the Hamiltonian becomes $\hat{H}_{\mathrm{ext}}=\hat{H}+\varepsilon_{{\o}}|{\o}\rangle\langle{\o}|$. $\varepsilon_{\o}$ is energy of the auxiliary state. For a spin chain with $S=\frac{1}{2}$ and $L_B=1$, we define a split operator as
\begin{equation}
  \hat{\mathcal{S}}=\Omega_{\uparrow}\hat{s}_{\uparrow}+\Omega_{\downarrow}\hat{s}_{\downarrow}.
\end{equation}
where $\Omega_{a}$ (with $a=\uparrow,\downarrow$) are arbitrary complex coefficients.
The operation of $\hat{\mathcal{S}}$ on a complete basis set in subsystem $B$ is specified as follows: $\hat{s}_{a}|a\rangle=|{\o}\rangle,\,
\hat{s}_{a}|\bar{a}\rangle=0,\,
\hat{s}_{a}|{\o}\rangle=0$.
Here, $|a\rangle$ are eigenstates of the Pauli matrix $\sigma^z$.
By definition, the split operator $\hat{\mathcal{S}}$ transfers a state from Hilbert space $\mathcal{H}$ to $\mathcal{H}_A\otimes\{|{\o}\rangle\}\otimes\mathcal{H}_B$ where the Hamiltonian $\hat{H}_{\mathrm{ext}}$ reduces to
\begin{equation}
  \hat{H}_{\mathrm{split}}=\hat{H}_A+\hat{H}_B+\varepsilon_{{\o}}|{\o}\rangle\langle{\o}|.
\end{equation}
Therefore, the operation of $\hat{\mathcal{S}}$ decouples subsystem $M$, effectively splitting the chain into $A$ and $B$; see Fig.\ref{fig1}. Although our analysis focuses on one-dimensional spin-1/2 systems throughout the paper, this framework can be readily extended to higher dimensions and more generic systems with larger spin values and subsystem sizes $L_M$.

We introduce split spectroscopy, defined as imaginary part of the Fourier transformation of Green's function
\begin{equation}\label{SplitSpec}
  \mathcal{A}(\omega)=-2\mathrm{Im}\int dt \mathcal{G}(t) e^{i\omega t},
\end{equation}
where the Green's function is given by
\begin{equation}\label{splitGF}
  \mathcal{G}(t)=-i\Theta(t)\langle \psi|\hat{\mathcal{S}}^\dag(t)\hat{\mathcal{S}}(0)|\psi\rangle.
\end{equation}
Here, $\hat{\mathcal{S}}^\dag(t)=e^{i\hat{H}_{\mathrm{ext}}t}\hat{\mathcal{S}}^\dag e^{-i\hat{H}_{\mathrm{ext}}t}$ is the split operator in Heisenberg picture, and $|\psi\rangle$ is an eigenstate of the Hamiltonian $\hat{H}$ with energy $\varepsilon$.
The spectral representation of the split spectroscopy can be derived as
\begin{equation}\label{SpecRep}
  \mathcal{A}(\omega)=\sum_{n=1}^{d_A}\sum_{m=1}^{d_B}|\gamma_{nm}
  |^2\delta(\omega+\varepsilon-\varepsilon^A_n-\varepsilon^B_m).
\end{equation}
Here, $\varepsilon_n^{X}$ is the energy of the n-th eigenstate $|\psi^{X}_n\rangle$ of the Hamiltonian $\hat{H}_X$, and $d_X$ denotes the dimensionality of the sub-Hilbert space $\mathcal{H}_X$. The coefficients $\gamma_{nm}$ represent overlap between the split state $\hat{\mathcal{S}}|\psi\rangle$ and eigenstates of the split chain $\hat{H}_{\mathrm{split}}$, $\gamma_{nm}=\langle{\o}|\langle \psi^A_n|\langle \psi^B_m|\hat{\mathcal{S}}|\psi\rangle$, see Appendix for details.

Under certain assumptions, we can establish the following criterion: \textit{The split spectroscopy exhibits a singe delta-function peak if and only if the investigated eigenstate is separable with respect to subsystems $A$, $B$ and $M$}; see Fig.\ref{fig1} and Appendix for the proof. Consequently, the split spectroscopy is both experimentally detectable and provides a necessary and sufficient criterion to verify the entanglement of eigenstates in quantum many-body systems, except in cases of complete degeneracy. Regarding spectroscopic measurements, a well-known phenomenon in conventional single-particle spectroscopies is that an uncorrelated system always exhibits a single delta-function peak spectrum \cite{Mahan@MBP}. Our criterion extends this observation to split spectroscopy for separable states. To quantitatively study spectrum, we introduce the spectral entropy given by
\begin{equation}
  E_{\mathrm{ent}}(|\psi\rangle)=-\sum_{nm}|\tilde{\gamma}_{nm}|^2\ln |\tilde{\gamma}_{nm}|^2
\end{equation}
with $|\tilde{\gamma}_{nm}|^2=\frac{|\gamma_{nm}|^2}{\sum_{nm}|\gamma_{nm}|^2}$. Up to a constant, it is equivalent to $-\int d\omega\tilde{\mathcal{A}}(\omega)\ln(\tilde{\mathcal{A}}(\omega))$
with normalized spectroscopy $\tilde{\mathcal{A}}(\omega)=\frac{\mathcal{A}(\omega)}{\int d\omega \mathcal{A}(\omega)}$.

\textit{The models.}
To implement our protocol for detecting entanglement through split spectroscopy, we consider two specific models: the XY model and the spin-1/2 Heisenberg chain in a random magnetic field. Their Hamiltonians are expressed as follows
\begin{equation}
\begin{aligned}
\hat{H}_{\mathrm{XY}}=&-\sum_{i=1}^{L-1}\left[J_x\hat{S}_i^x \hat{S}_{i+1}^x+J_y \hat{S}_i^y \hat{S}_{i+1}^y\right]+h \sum_{i=1}^L \hat{S}_i^z,\\
\hat{H}_{\mathrm{RF}}=&J\sum_{i=1}^{L-1}\hat{\mathbf{S}}_i\cdot\hat{\mathbf{S}}_{i+1}+\sum_{i=1}^L h_i\hat{S}_i^z.
\end{aligned}
\end{equation}
Here, $\hat{S}_i^{\alpha=x,y,z}$ represents the standard spin-1/2 operator located at site $i$. In the XY model, the nearest neighbour couplings are given by $J_x=J(1+\alpha),J_y=J(1-\alpha)$ where $\alpha$ quantifies the anisotropy in the in-plane interaction. This study focuses on the cases where $\alpha=0$ and $\alpha=1$ \cite{Sachdev@QPT}. For $\alpha=0$, it simplifies to the isotropic XY model, wherein the energy gap vanishes for transverse field values $h/J<1$, indicating a quantum phase transition occurs at $h_c/J=1$. For $\alpha=1$, it becomes transverse Ising model, which also exhibits a quantum phase transition at critical value $h_c=1$. In the random field model, we sample random fields $h_i$ from a uniform distribution ranging over $[-H,H]$. The system is believed to display a many-body localization (MBL) transition at $H_c/J\approx3.6$ \cite{Huse@2010}. The eigenstates of these models across different phases exhibit varying levels of entanglement \cite{Vidmar@2022,Plenio@2010}, thereby enabling a verification of our protocol.

\begin{figure}[tbp]
  \centering
  \includegraphics[width=\columnwidth]{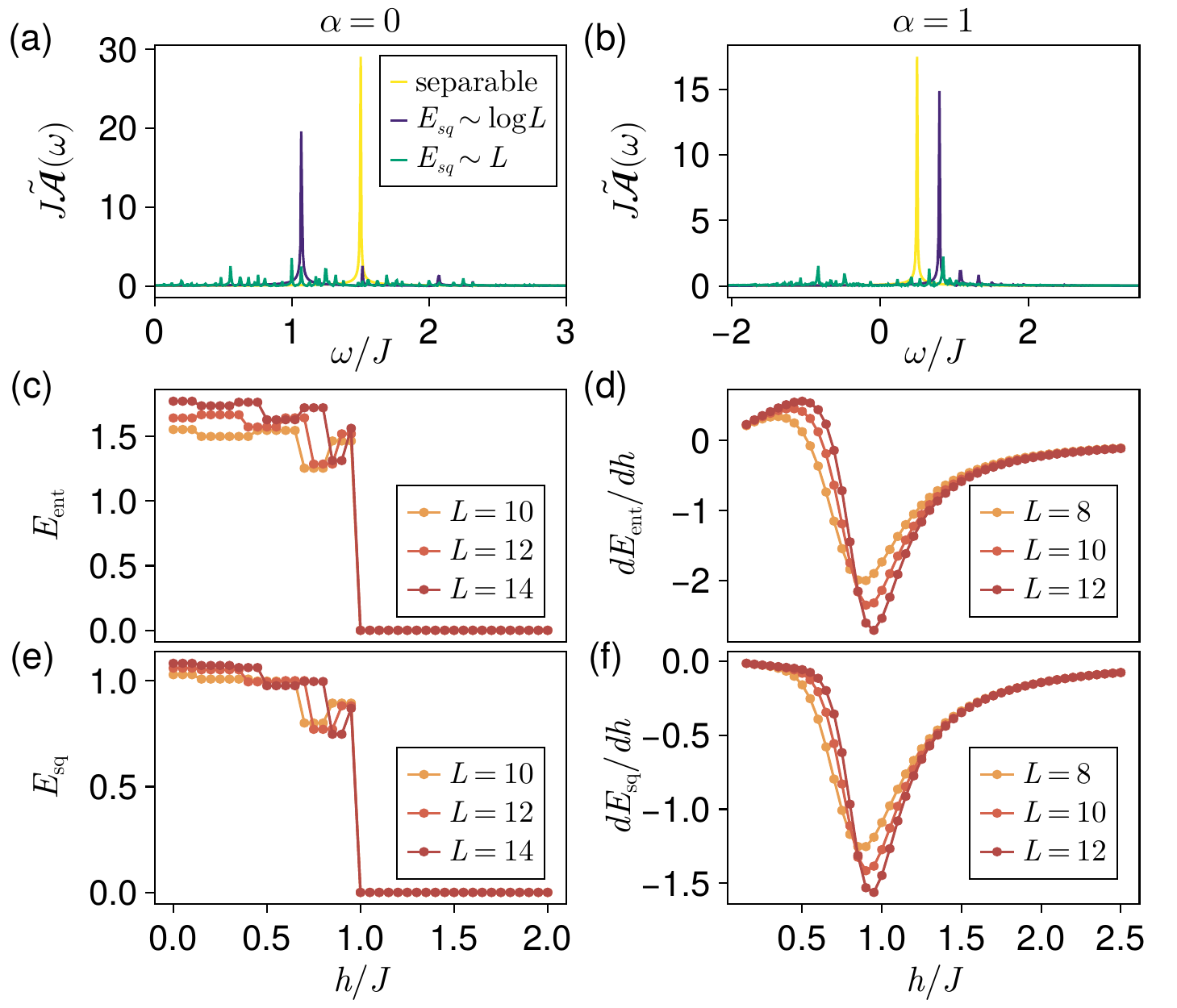}
  \caption{Numerical results for the XY model. (a)-(b) The normalized coefficients $|\tilde{\gamma}_{nm}|^2$. The yellow line corresponds to the separable ground state at $h/J=1.5$ for $\alpha=0$, and $J=0,h=1$ for $\alpha=1$. The purple line corresponds to ground state with logarithm-law entanglement at gapless phase $h/J=0$ for $\alpha=0$ and $h/J=1$ for $\alpha=1$, while the green line denotes a typical excited state with volume-law entanglement. (c)-(f) The spectral entropy and squashed entanglement of ground state as a function of the field strength in different system size.}\label{fig2}
\end{figure}

\textit{Numerical results.}
For comparison, we also calculate a standard entanglement measure---the multipartite squashed entanglement \cite{Song@2009}, which quantifies the entanglement among subsystems $A$, $B$, and $M$. Focusing on pure states, the squashed entanglement can be expressed as the simplified form
\begin{equation}
  E_{\mathrm{sq}}=\frac{1}{2}\left[S(A)+S(B)+S(M)\right].
\end{equation}
Here, $S(X)$ is the von Neumann entropy of subsystem $X$. In this paper, we designate the subsystem $M$ as a single site located at the center of the chain.

\begin{figure}[tbp]
  \centering
  \includegraphics[width=\columnwidth]{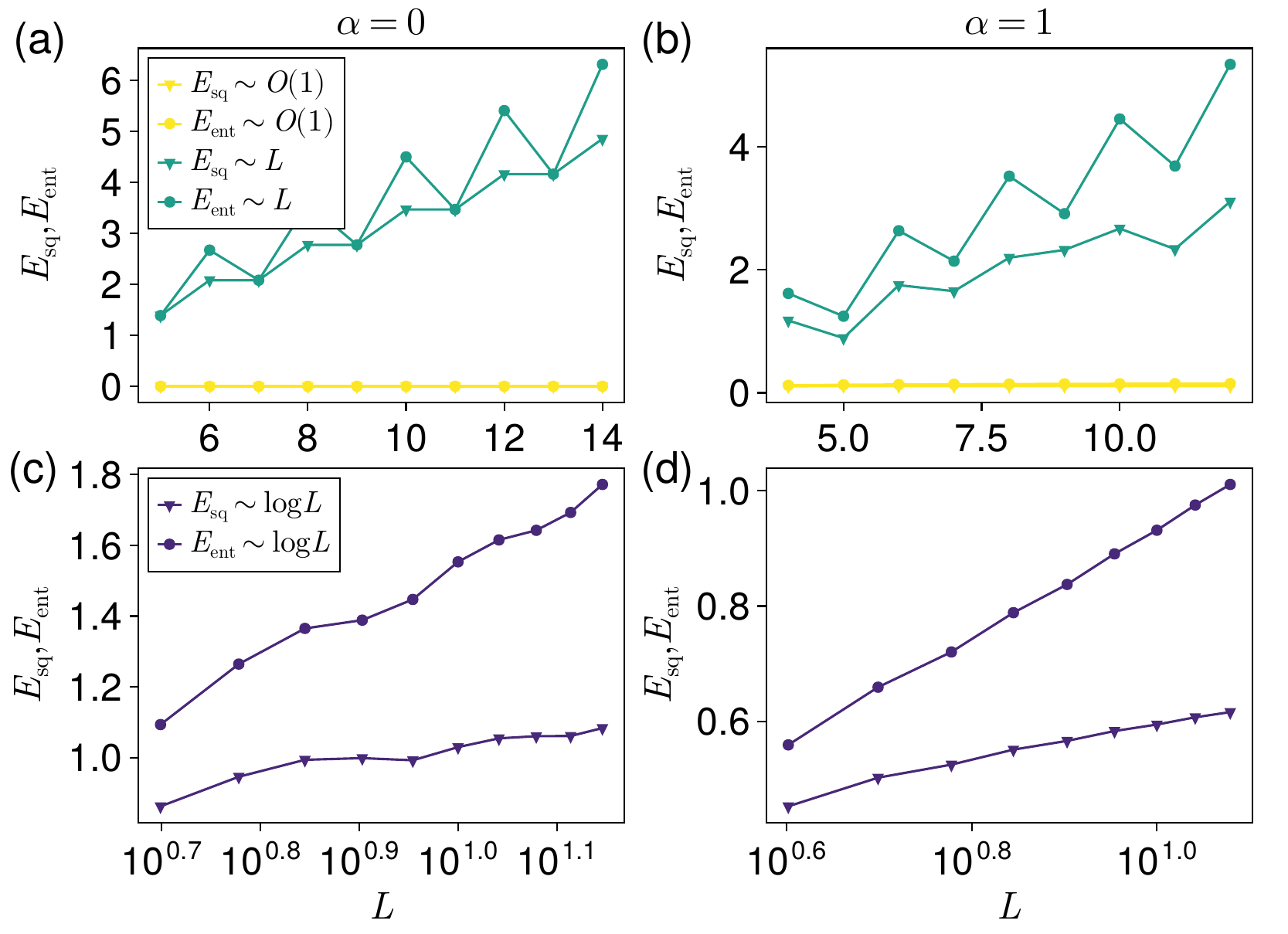}
  \caption{Comparison of the scaling behavior between spectral entropy $E_{\mathrm{ent}}$ and squashed entanglement $E_{\mathrm{sq}}$ for $\alpha=0$ (a),(c) and $\alpha=1$ (b),(d). (a)-(b) In gapped phases $\{\alpha,J,h\}=\{0,1,1.5\}$ and $\{\alpha,J,h\}=\{1,1,2.5\}$, the ground states exhibit area-law entanglement (yellow line), while typical excited states display volume-law entanglement (green line). (c)-(d) In gapless regions $\{\alpha,J,h\}=\{0,1,0\}$ and $\{\alpha,J,h\}=\{1,1,1\}$, the entanglement of the ground states scales with the logarithm of system size. The spectral entropy shares the same scaling behavior as squashed entanglement.}\label{fig3}
\end{figure}

Firstly, we numerically investigate the XY chain with parameters $\{\alpha,J,h\}=\{0,1,1.5\}$ and $\{\alpha,J,h\}=\{1,0,1\}$, where the ground state is a product state of spins,  $|\psi_{gs}\rangle=|\downarrow,\cdots,\downarrow\rangle$. We computer the coefficients $|\gamma_{nm}|^2$  associated with $|\psi_{gs}\rangle$ in the Green's function Eq.\eqref{GF}, Subsequently, we obtain the spectroscopy $\mathcal{A}(\omega)$ via Fourier transformation. The results are shown in Fig.\ref{fig2}(a)-(b). We find only one nonzero coefficients $|\gamma_{nm}|^2=1$, resulting in the spectroscopy displaying a single delta peak. By definition, the corresponding spectral entropy $E_{\mathrm{ent}}(|\psi_{gs}\rangle)$ is zero. For parameter sets $\{\alpha,J,h\}=\{0,1,0\}$ and $\{\alpha,J,h\}=\{1,1,1\}$, both the ground states and excited states become non-separable regarding subsystems A, B and M. The squashed entanglement of the ground state scales logarithmically with the system size, and the excited state exhibits volume-law scaling, as shown in Fig.\ref{fig3}. In this case, the spectroscopy exhibits multiple delta-function peaks, leading to a non-zero spectral entropy, shown in Fig.\ref{fig2}. Additionally, we also obtain similar results for different parameters and in various models, including the XXZ model and the Majumdar-Ghosh model.
These findings further substantiate that our protocol offers a reliable method for detecting entanglement in experiments.

\begin{figure}[tbp]
  \centering
  \includegraphics[width=\columnwidth]{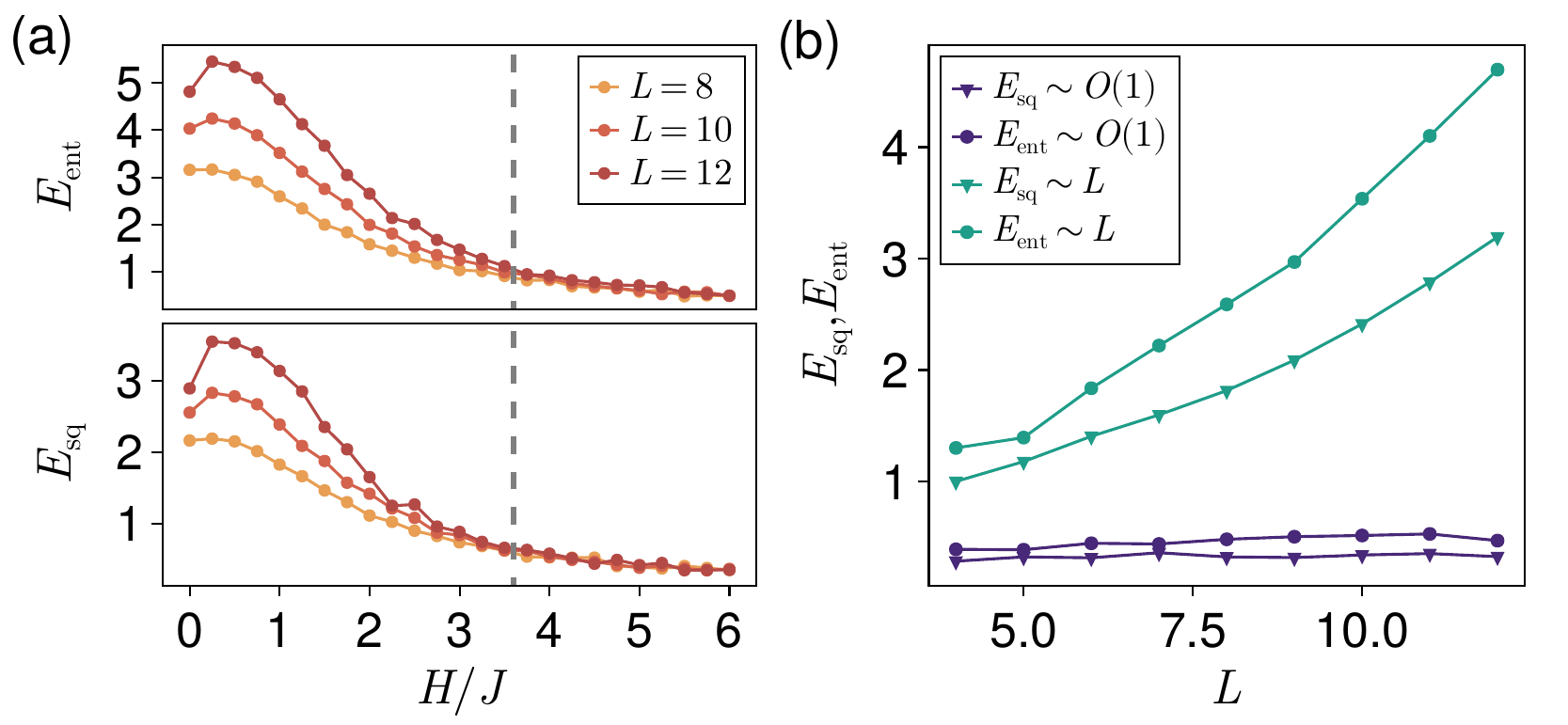}
  \caption{Numerical results for the random field model. (a) The average spectral entropy and squashed entanglement for various values of $H$ in different system size. Their scaling behavior has a transition near the critical point $H=3.6$ marked by the gray vertical line. (b) The scaling of the spectral entropy and squashed entanglement display a volume law in the normal phase $H=1$ and area law in the MBL phase $H=5$.}\label{fig4}
\end{figure}

Similar to other entanglement measures, the split spectrum can also indicate quantum phase transitions. We investigate the behavior of spectral entropy near quantum critical points. Figs \ref{fig2}(c)(e) present the spectral entropy $E_{\mathrm{ent}}$ and squashed entanglement $E_{\mathrm{sq}}$ of groundstate for $\alpha=0$, while Figs \ref{fig2}(d)(f) show their derivatives for $\alpha=1$. For $\alpha=0$, the spectral entropy is zero in phase with $h>h_c$ because the ground state is a product state. It exhibits finite values that increases with system size for the entangled ground state in phase with $h<h_c$. At the critical point $h_c=1$, a sudden change occurs. For $\alpha=1$, the derivative $dE_{\mathrm{ent}}/dh$ diverges at the critical point as the system approaches the thermodynamic limit. In the random field model, we numerically calculate the average spectral entropy over at least $100$ highly excited eigenstates and $100$ disorder realizations for various values of $H$. We find the scaling of $E_{\mathrm{ent}}$ undergoes a sudden change near critical point $H/J\approx3.6$, as illustrated in Fig.\ref{fig4}(a). These critical behaviors are analogous to those of squashed entanglement, and align with other established results \cite{Meng@2025,Huse@2010}.

We further analyze the scaling behavior of spectral entropy and compare it with that of squashed entanglement. In the XY model, the spectral entropy of ground state in gapped phase scales with the area of system,  $E_{\mathrm{ent}}(|\psi_{gs}\rangle)\sim O(1)$, as shown in Fig.\ref{fig3}. In contrast, at critical points (gapless regions), it scales with the logarithm of the system size, $E_{\mathrm{ent}}(|\psi_{gs}\rangle)\sim \log L$. The spectral entropy of typical excited states scales with the volume of system, $E_{\mathrm{ent}}(|\psi_{es}\rangle)\sim L$. In the random field model, the average spectral entropy exhibits volume law in normal phase and area law in MBL phase, as illustrated in Fig.\ref{fig4}(b). Notably, we find that spectral entropy shares the same scaling behavior as squashed entanglement. This indicates the split spectroscopy can not only verify entanglement but also capture its scaling properties, which have attracted significant attention \cite{Bousso@2002,Rigol@2018,Srednicki@2019,Vidmar@2022,Kitaev@2003,Plenio@2010}.

\textit{Experimental realization in Rydberg atom arrays.}
Recently, Rydberg atom arrays have emerged as a new platform for implementing quantum computation \cite{Jurczak@2020} and simulating quantum many-body systems \cite{Lukin@2021,Zhai@2024}, in which an effective spin-1/2 is encoded in a pair of internal states of the Rydberg atom. Here, we choose the ground sate $|g\rangle$ and Rydberg state $|R\rangle$. To perform the split spectroscopy measurement, we employ two radio frequency waves at the same frequency $\omega_{1}=\omega_2=\omega$, which transfer atoms in states $|g\rangle$ and $|R\rangle$ to a third internal state $|i\rangle$ respectively. The state $|i\rangle$ is out of the investigated spin system. The light-matter interaction Hamiltonian is given by
\begin{equation}\label{light-matter}
\begin{aligned}
  \hat{H}_I(t)=&\sum_{r} e^{-i\omega t}\hat{\psi}^\dag_{ri}\left(\Omega_1\hat{\psi}_{rg}+\Omega_2\hat{\psi}_{rR}\right)+\mathrm{h}.\mathrm{c}.\\
  =&\sum_{r} e^{-i\omega t}\hat{\psi}^\dag_{ri}\hat{\mathcal{S}}_{r}+\mathrm{h}.\mathrm{c}.
\end{aligned}
\end{equation}
where $\hat{\psi}_{r\nu}$ ($\nu=g,R,i$) represents the annihilation operator for an atom in state $|\nu\rangle$ located at position $r$, and $\Omega_{1,2}$ are the Rabi frequencies for transitions from $|g\rangle$ and $|R\rangle$ to $|i\rangle$. We have introduced the split operator at position $r$ as $\hat{\mathcal{S}}_{r}=\Omega_1\hat{\psi}_{rg}+\Omega_2\hat{\psi}_{rR}$. The Hamiltonian $\hat{H}_I(t)$ effectively describes a single radio frequency wave process, and has been extensively studied \cite{Shen@2021,Levin@2009}. The first-order response of the transition rate $\frac{d\langle \hat{n}_{ir}\rangle(t)}{dt}$ to the external perturbation $\hat{H}_I(t)$ gives the split spectroscopy
\begin{equation}
  \frac{d\langle \hat{n}_{ir}\rangle(t)}{dt} = -2\mathrm{Im}\int\mathrm{d}t \mathcal{G}(t)e^{i\omega t},
\end{equation}
where $\hat{n}_{ir}=\hat{\psi}^\dag_{ri}\hat{\psi}_{ri}$ is the number operator for the third state $|i\rangle$, and the Green's function is provided in Eq.\eqref{splitGF}.

\textit{Discussion.}
In summary, we introduce the split spectroscopy as a scalable experimental method for detecting
entanglement in many-body systems. Under certain assumptions, we establish an necessary and sufficient criterion. To illustrate our framework, we study two one-dimensional quantum spin models. Numerical results indicate that the spectral entropy, analogous to squashed entanglement, is a powerful tool for identifying quantum critical points and captures the scaling behavior of entanglement. Furthermore, the extension to more generic fermionic or bosonic systems is straightforward. In higher dimensions, split spectroscopy could be realized by applying multiple RF waves to split the system, or directly detect the single-site entanglement \cite{Zanardi@2002,Lin@2004,Ortolani@2006}.

We conclude with several open questions and future directions. Our study focuses on detecting entanglement of eigenstates in quantum systems. Thus, it is meaningful to extend this formalism to Gibbs thermal states. In spectroscopic measurements, the delta-function peak for separable states is an analogue to the delta-function behavior observed in single-particle spectroscopies for uncorrelated systems \cite{Mahan@MBP}. A natural question arises: is there an analytic formalism to characterize the split spectroscopy for entangled states, similar to the diagrammatic perturbation theory for correlated systems. We defer a careful analysis of these questions to future work.

\textit{Acknowledgments.}
This work is supported by NSFC (Grants No. GG2030040453 and No. GG2030007011) and the Innovation Program for Quantum Science and Technology (Grant No. 2021ZD0302004).


\newpage
\begin{widetext}
\appendix

\section{Spectral representation}
Now, let us derive the spectral representation of the split spectroscopy. In general, an eigenstate $|\psi\rangle$ can be expended as
\begin{equation}
  |\psi\rangle=\sum_{n=1}^{d_A}\sum_{l=1}^{d_M}\sum_{m=1}^{d_B}\alpha_{nlm}|\psi^A_n\rangle|\psi^M_l\rangle|\psi^B_m\rangle,
\end{equation}
where $|\psi^{X}_n\rangle$ is the $n$-th eigenstate of Hamiltonian $H_{X}$ with energy $\varepsilon^{X}_n$, and $d_X$ is dimension of the sub-Hilbert space $\mathcal{H}_X$. For a spin chain with $S=\frac{1}{2}$ and $L_M=1$, apply split operator $\hat{\mathcal{S}}$ to the state $|\psi\rangle$ to obtain
\begin{equation}\label{SE}
  \hat{\mathcal{S}}|\psi\rangle=\sum_{n=1}^{d_A}\sum_{m=1}^{d_B}\gamma_{nm}|\psi^A_n\rangle|\psi^B_m\rangle|{\o}\rangle,
\end{equation}
where $\gamma_{nm}=\sum_{l=1}^{d_M}\alpha_{nlm}
\left(\langle\uparrow|\psi^M_l\rangle\Omega_\uparrow+\langle\downarrow|\psi^M_l\rangle\Omega_\downarrow\right)$
represents the overlap between $\hat{\mathcal{S}}|\psi\rangle$ and the eigenstates of $\hat{H}_{\mathrm{split}}$. Substituting Eq.\eqref{SE} into Green's function, and using $\hat{H}_{M,AM,MB}|\o\rangle=0$ yield
\begin{equation}\label{GF}
  \mathcal{G}(t)=-i\Theta(t)\sum_{n=1}^{d_A}\sum_{m=1}^{d_B}
  |\gamma_{nm}|^2e^{-i(\varepsilon^A_n+\varepsilon^B_m-\varepsilon)t}.
\end{equation}
Here we have set $\varepsilon_{{\o}}=0$ for convenience. Ultimately, the spectral representation is given by
\begin{equation}
  \mathcal{A}(\omega)=\sum_{n=1}^{d_A}\sum_{m=1}^{d_B}
  |\gamma_{nm}|^2\delta(\omega+\varepsilon-\varepsilon^A_n-\varepsilon^B_m).
\end{equation}

\section{Proof of the criterion}
In this section, we prove the criterion: the split spectroscopy exhibits a singe delta-function peak if and only if the investigated eigenstate is separable with respect to subsystems A, B and M. Given the spectral representation and the meaning of coefficients $\gamma_{nm}$, it suffices to demonstrate the following two criterions.

\textit{criterion 1}: If the eigenstate $|\psi\rangle$ of $\hat{H}$ is separable with respect to subsystems $A,B$ and $M$, i.e., $|\psi\rangle=|\phi^A\rangle|\phi^M\rangle|\phi^B\rangle$, then the split state $\hat{\mathcal{S}}|\psi\rangle$ is an eigenstate of $\hat{H}_{\mathrm{split}}$.

\textbf{Proof:} Firstly, we consider a bipartite separable case, described by
\begin{equation}\label{condition}
  (\hat{H}_A+\hat{H}_M+\hat{H}_{AM})|\phi^A\rangle|\phi^M\rangle=\varepsilon|\phi^A\rangle|\phi^M\rangle.
\end{equation}
This condition implies that $\hat{H}_{AM}$ does not generate entanglement between subsystems $A$ and $M$, and its action on the separable state can be expressed as the generic form
\begin{equation}\label{conditionHAB}
  \hat{H}_{AM}|\phi^A\rangle|\phi^M\rangle=(\lambda_A|\phi^A\rangle+\beta_A|\phi^{A}_{\perp}\rangle)
  (\lambda_M|\phi^M\rangle+\beta_M|\phi^{M}_{\perp}\rangle)
\end{equation}
with $\langle\phi^X|\phi^X_{\perp}\rangle=0$, $X=A,M$. Substituting Eq.\eqref{conditionHAB} into Eq.\eqref{condition}, we deduce that at least one of $\beta_A$ and $\beta_M$ must vanish.

If $\beta_A\ne0, \beta_M=0$, the Hamiltonian $\hat{H}_A$ must be fine-tuned to satisfy
\begin{equation}
  \hat{H}_A|\phi^A\rangle=\varepsilon^A|\phi^A\rangle-\beta_A\lambda_M|\phi^{A}_{\perp}\rangle
\end{equation}
precisely canceling the term $\beta_A\lambda_M|\phi^{A}_{\perp}\rangle$ in Eq.\eqref{conditionHAB}.
Such fine-tuning is generically impossible in physical systems. A similar argument holds for the case $\beta_M\ne0, \beta_A=0$. Thus, the only remaining case is $\beta_A=\beta_M=0$, leading to
\begin{equation}
  (\hat{H}_A+\hat{H}_M)|\phi^A\rangle|\phi^M\rangle=(\varepsilon-\lambda_A\lambda_M)|\phi^A\rangle|\phi^M\rangle.
\end{equation}
This further implies that $|\phi^{A,M}\rangle$ is an eigenstate of $\hat{H}_{A,M}$, specifically,
\begin{equation}
  \hat{H}_{A,M}|\phi^{A,M}\rangle=\varepsilon^{A,M}|\phi^{A,M}\rangle.
\end{equation}

For the tripartite separable case, we can also obtain $\hat{H}_B|\phi^{B}\rangle=\varepsilon^{B}|\phi^{B}\rangle$ by applying the above analysis twice. Therefore, $\hat{\mathcal{S}}|\psi\rangle\propto|\phi^A\rangle|{\o}\rangle|\phi^B\rangle$ is an eigenstate of $\hat{H}_{\mathrm{split}}$.

\textit{criterion 2:} If the eigenstate $|\psi\rangle$ of $\hat{H}$ is  nonseparable with respect to subsystems $A,B$ and $M$, then the split state $\hat{\mathcal{S}}|\psi\rangle$ exhibits overlap with multiple eigenstates of $\hat{H}_{\mathrm{split}}$.

\textbf{Proof:} Without loss of generality, assume that entanglement exists between subsystems $A$ and $M$, while subsystem $B$ is separable. Then, $|\psi\rangle$ can be expanded as
\begin{equation}
  |\psi\rangle=\sum_{n}\sum_{l}\alpha_{nl}|\varepsilon^A_{n}\rangle|\varepsilon^M_{l}\rangle|\varepsilon^B\rangle.
\end{equation}
where $|\varepsilon^X_n\rangle$ is the n-th eigenstate of $\hat{H}_X$. Applying split operator on state $|\psi\rangle$ we obtain
\begin{equation}
  \hat{\mathcal{S}}|\psi\rangle=\sum_{n}\gamma_n|\varepsilon^A_n\rangle|\varepsilon^B\rangle|{\o}\rangle
\end{equation}
with $\gamma_{n}=\sum_{l=1}\alpha_{nl}
\left(\langle\uparrow|\varepsilon^M_l\rangle\Omega_\uparrow+\langle\downarrow|
\varepsilon^M_l\rangle\Omega_\downarrow\right)$.
For a generic entangled state, $\gamma_n$ is non-zero in at least two summation terms.
Therefore, multiple eigenstates of $\hat{H}_{\mathrm{split}}$ exhibit overlap with $\hat{\mathcal{S}}|\psi\rangle$. Here, we assume that eigenstate $|\varepsilon^A_n\rangle$ with non-zero coefficient $\gamma_n$ are not entirely degenerate, ensuring that their spectral peaks remain distinct.
\end{widetext}

\end{document}